\documentclass{article}
\usepackage{dcase2023,amsmath,graphicx,url,times,booktabs, tabularx}

\usepackage{amsmath,comment}
\usepackage{
amsfonts,
mathrsfs,
url,
xurl,
cite,
booktabs,
tabularx,
array,
multirow,
}

\newlength{\tablelength}
\newcolumntype{C}[1]{>{\centering\arraybackslash}p{#1\tablelength}}

\newcommand{\mysection}[1]{
    \vspace{-2pt}
    \section{#1}
    \vspace{-3pt}
    }
    
\newcommand{\mysubsection}[1]{
    \vspace{-3pt}
    \subsection{#1}
    \vspace{-3pt}
    }

\newcommand{\myqt}[1]{\textit{#1}}
\title{Audio Difference Captioning \\ Utilizing Similarity-Discrepancy Disentanglement}
\name{Daiki Takeuchi, Yasunori Ohishi, Daisuke Niizumi, Noboru Harada, and Kunio Kashino}
\address{
NTT Corporation, Japan
}

\begin{document}
\sloppy

\maketitle
\ninept 
\begin{abstract}
We proposed \textit{Audio Difference Captioning} (ADC) as a new extension task of audio captioning for describing the semantic differences between input pairs of similar but slightly different audio clips.
The ADC solves the problem that conventional audio captioning sometimes generates similar captions for similar audio clips, failing to describe the difference in content.
We also propose a cross-attention-concentrated transformer encoder to extract differences by comparing a pair of audio clips and a similarity-discrepancy disentanglement to emphasize the difference in the latent space.
To evaluate the proposed methods, we built an AudioDiffCaps dataset consisting of pairs of similar but slightly different audio clips with human-annotated descriptions of their differences.
The experiment with the AudioDiffCaps dataset showed that the proposed methods solve the ADC task effectively and improve the attention weights to extract the difference by visualizing them in the transformer encoder.

\end{abstract}

\section{Introduction}
\label{sec:intro} 

\begin{figure}[t]
  \centering
\includegraphics[width=0.9\columnwidth]{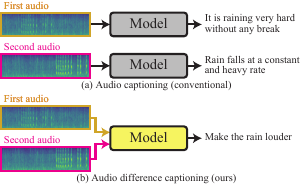} 
\vspace{-5pt}
  \caption{
  Conceptual diagram of conventional audio captioning and audio difference captioning.
  Audio difference captioning describes the difference between pair audio clips, while conventional audio captioning describes the contents of each. 
  }
  \label{fig:task_illust}
  \vspace{-10pt}
\end{figure}

Audio captioning is used to generate the caption for an audio clip~\cite{kim2019audiocaps, drossos2019clotho, takeuchi2020effects, xu2020crnn, mei2021audio, gontier2021automated, koizumi2020audio, xu2022diversity, mei2022diverse, liu2022leveraging}.
Unlike labels for scenes and events\cite{piczak2015dataset, barchiesi2015acoustic, mesaros2016tut, gemmeke2017audio, fonseca2020fsd50k}, captions describe the content of the audio clip in detail.
However, conventional audio captioning systems often produce similar captions for similar audio clips, making it challenging to discern their differences solely based on the generated captions.
For instance, suppose two audio clips of heavy rain are input into a conventional captioning system. 
The system will generate a caption describing the content of each, like ``\textit{It is raining very hard without any break}'' and ``\textit{Rain falls at a constant and heavy rate}''\footnote{These captions were taken from the Clotho dataset~\cite{drossos2019clotho}} as illustrated in Fig.~\ref{fig:task_illust}(a). 
The difference, such as which rain sound is louder, is difficult to understand from the generated captions in this case.

To address this problem, we propose \textit{Audio Difference Captioning} (ADC) as a new extension task of audio captioning. 
ADC takes two audio clips as input and outputs text explaining the difference between two inputs as shown in Fig.~\ref{fig:task_illust}. 
We make the ADC clearly describe the difference between the two audio clips,
such as ``Make the rain louder,'' which describes what and how to modify one audio clip to the other in the instruction form, even for audio clips with similar texts. 
Potential real-world applications include machine condition and healthcare monitoring using sound by captioning anomalies that differ from usual sounds.

The ADC task has two major challenges: different content detection and detection sensitivity. 
Since the difference between a pair of audio clips can be classes of contained events or an attribute, such as loudness, the ADC needs to detect what difference to describe. 
When the difference lies in an attribute, the ADC needs to be sensitive enough to detect the magnitude of the attribute, such as rain is hard or moderately shown in the example in Fig.~\ref{fig:task_illust}.

To handle these challenges, the ADC should extract features of difference based on the cross-reference of two audio clips. 
These features should carry enough information to differentiate critical attributes such as loudness. 
A typical choice of a feature extractor could be pre-trained models to classify labels~\cite{hershey2017cnn, kong2020panns, gong2021ast}. However, these models learn to discriminate sound event classes, learning what is common while ignoring subtle differences such as raining hard or quietly unless the class definition covers that.

To meet the requirements of the ADC mentioned above, we propose (I) a cross-attention-concentrated~(CAC) transformer encoder and (II) a similarity-discrepancy disentanglement~(SDD). 
The CAC transformer encoder utilizes the masked multi-head attention layer, which only considers the cross-attention of two audio clips to extract features of difference efficiently. 
The SDD emphasizes the difference feature in the latent space using contrastive learning based on the assumption that two similar audio clips consist of similar and discrepant parts.

We demonstrate the effectiveness of our proposals using a newly built dataset, AudioDiffCaps, consisting of two similar but slightly different audio clips synthesized from existing environmental sound datasets~\cite{fonseca2020fsd50k, piczak2015dataset} and human-annotated difference descriptions. 
Experiments show that the CAC transformer encoder improves the evaluation metric scores by making the attention focus only on cross-references. The SDD also improves the scores by emphasizing the differences between audio clips in the latent space. 
Our contributions are proposals of (i) the ADC task, (ii) the CAC transformer encoder and SDD for solving ADC, (iii) the AudioDiffCaps dataset, and (iv) demonstrating the effectiveness of these proposals.

\begin{figure*}[t]
\vspace{-9pt}
  \centering
\includegraphics[width=1.95\columnwidth]{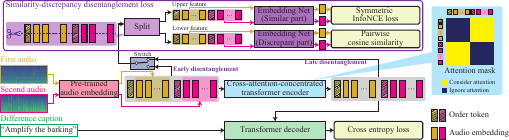} 
\vspace{-5pt}
  \caption{
  Model architecture of our proposed method. 
  The cross-attention-concentrated transformer encoder uses an attention mask illustrated in the upper left. 
  The similarity-discrepancy disentanglement is conducted by symmetric InfoNCE loss and pairwise cosine similarity. The input to them is either the input or output of the cross-attention-concentrated transformer encoder.}
  \label{fig:model_arch}
  \vspace{-10pt}
\end{figure*}

\section{Audio difference captioning}
We propose ADC, a task for generating texts to describe the difference between two audio clips.
ADC estimates a word sequence $\boldsymbol{w}$ from the two audio clips $\boldsymbol{x}$ and $\boldsymbol{y}$. 

The general framework to solve ADC includes three main functions: audio embedding, audio difference encoding, and text decoding.
Audio embedding calculates two audio embedding vectors from two audio clips, respectively.
Audio difference encoding captures the difference between two audio embedding vectors.
Text decoding generates a description of the differences from captured differences.
Audio embedding and audio difference encoding require approaches specific to ADC.
In particular, difference encoding is the function unique to audio difference captioning.
This function requires a model structure to capture the subtle differences between two audio clips, unlike conventional audio captioning that captures the content of a single audio clip.
Moreover, the sensitivity to the subtle difference between two similar audio clips is also necessary for audio embedding.
The pre-trained audio embedding models widely used for conventional environmental sound analysis tasks are often trained for classification tasks and are suitable for identifying predefined labels.
Consequently, the outputs of these pre-trained audio embedding models are not sensitive to the subtle differences between audio clips with the same label.
Therefore, learning to emphasize the differences between similar audio clips in the latent space is necessary when applying pre-trained audio embedding models to the ADC.

\mysection{Proposed method}
Based on the above discussion, we propose the ADC system illustrated in Fig.~\ref{fig:model_arch}.
Our system consists of an audio feature extractor~(red), difference encoder~(blue), text decoder~(green), and similarity-discrepancy disentanglement~(purple).

\mysubsection{Audio feature extractor}
The audio feature extractor uses a pre-trained audio embedding model to calculate audio embedding vectors.
Two audio clips $\boldsymbol{x}$ and $\boldsymbol{y}$ are the input, and the audio embedding vectors corresponding to the clips $\boldsymbol{X} \in \mathbb{R}^{H \times T_x}$ and $\boldsymbol{Y} \in \mathbb{R}^{H \times T_y}$ are the output, where $H$ is the size of hidden dimension, $T_x$ is the time length of $\boldsymbol{X}$, and $T_y$ is the time length of $\boldsymbol{Y}$

\mysubsection{Difference encoder}
The difference encoder extracts information about the differences between the two audio clips from audio embedding vectors $\boldsymbol{X}$ and $\boldsymbol{Y}$.
To extract difference information efficiently, we utilize a cross-attention-concentrated~(CAC) transformer encoder as the main function of the difference encoder.
The CAC transformer encoder utilizes the masked multi-head attention layer, allowing only mutual cross-attention between two audio clips by the attention mask illustrated in the upper right of Fig.~\ref{fig:model_arch}.

The detailed procedure is as follows.
First, special tokens that indicate the order of the audio clips $\mathcal{X} \in \mathbb{R}^{H \times 1}$ and $\mathcal{Y} \in \mathbb{R}^{H \times 1}$ are concatenated at the beginning of $\boldsymbol{X}$ and $\boldsymbol{Y}$, respectively.
Next, these two sequences are concatenated to make the input of the difference encoder $\boldsymbol{Z}$ like $\boldsymbol{Z} = [\mathcal{X}, \boldsymbol{X}, \mathcal{Y}, \boldsymbol{Y}]$.
Then, positional encoding $\mathscr{P}$ is applied to $\boldsymbol{Z}$.
Finally, $\mathscr{P}(\boldsymbol{Z})$ is input to CAC transformer encoder to obtain the output $\boldsymbol{\hat{Z}}= [\hat{\mathcal{X}}, \boldsymbol{\hat{X}}, \hat{\mathcal{Y}}, \boldsymbol{\hat{Y}}]$.

\mysubsection{Text decoder}
The transformer decoder is utilized as a text decoder like as \cite{mei2021audio}.
The text decoder calculates word probability from the output of the difference encoder $\boldsymbol{\hat{Z}}$.

\mysubsection{Similarity-discrepancy disentanglement}
The similarity-discrepancy disentanglement~(SDD) loss function is an auxiliary loss function aimed at obtaining a difference-emphasized audio representation.
When there is an explainable difference between two audio clips, these clips consist of similar and discrepant parts.
To introduce this hypothesis, we design contrastive learning to bring similar parts closer and keep discrepant parts.
We propose two types of implementations that apply SDD to the input of the difference encoder $\boldsymbol{Z}$ or the output of it $\boldsymbol{\hat{Z}}$, as shown in Fig.~\ref{fig:model_arch}, and call the former and latter implementations early and late disentanglement, respectively.

We explain the procedure in the case of early disentanglement.
Note that the case of late disentanglement only replaces  $\boldsymbol{Z}$ with $\boldsymbol{\hat{Z}}$.
First, $\boldsymbol{Z}$ is split along the hidden dimension and assigned to similar and discrepant parts like in the upper left illustration of Fig.~\ref{fig:model_arch}.
If $\boldsymbol{Z} \in \mathbb{R}^{H \times (T_x+T_y+2)}$, $\boldsymbol{Z}$ is split into similar part $\boldsymbol{Z}_{\rm S}$ and discrepant part $\boldsymbol{Z}_{\rm D}$ like
\begin{align}
    \boldsymbol{Z}_{\rm S} &= [\mathcal{X}_{\rm S}, \boldsymbol{X}_{\rm S}, \mathcal{Y}_{\rm S}, \boldsymbol{Y}_{\rm S}] \in \mathbb{R}^{(H/2) \times (T_x+T_y+2)}, \\
    \boldsymbol{Z}_{\rm D} &= [\mathcal{X}_{\rm D}, \boldsymbol{X}_{\rm D}, \mathcal{Y}_{\rm D}, \boldsymbol{Y}_{\rm D}] \in \mathbb{R}^{(H/2) \times (T_x+T_y+2)}.
\end{align}
Then, the SDD is performed by $\mathcal{L}_{\rm SDD} = \mathcal{L}_{\rm S} + \mathcal{L}_{\rm D}$, where
\begin{align}
    &\mathcal{L}_{\rm S} = {\rm SymInfoNCE}(\Phi([\mathcal{X}_{\rm S}, \boldsymbol{X}_{\rm S}]), \Phi([\mathcal{Y}_{\rm S}, \boldsymbol{Y}_{\rm S}])), \\
    &\mathcal{L}_{\rm D} = {\rm PairCosSim}(\Psi([\mathcal{X}_{\rm D}, \boldsymbol{X}_{\rm D}]), \Psi([\mathcal{Y}_{\rm D}, \boldsymbol{Y}_{\rm D}])), 
\end{align}
${\rm SymInfoNCE}$ is the symmetric version of the InfoNCE loss used in~\cite{radford2021learning}, ${\rm PairCosSim}$ is the cosine similarity for each correct data pair,  $\Phi$ and $\Psi$ are embedding networks consisting of the bidirectional-LSTM and average pooling, and $\mathcal{L}_{\rm SDD}$ is the final value of the SDD loss function.
That is, the SDD loss function views $[\mathcal{X}_{\rm S}, \boldsymbol{X}_{\rm S}]$ and $[\mathcal{Y}_{\rm S}, \boldsymbol{Y}_{\rm S}]$ as similar parts and brings them closer by using $\mathcal{L}_{\rm S}$ and views $[\mathcal{X}_{\rm D}, \boldsymbol{X}_{\rm D}]$ and $[\mathcal{Y}_{\rm D}, \boldsymbol{Y}_{\rm D}]$ as discrepant parts and keeps them apart by $\mathcal{L}_{\rm D}$.

The entire loss function $\mathcal{L}$ is the weighted sum of cross-entropy loss for word prediction $\mathcal{L}_{\rm CE}$ and the SDD: $\mathcal{L} = \mathcal{L}_{\rm CE} + \lambda \mathcal{L}_{\rm SDD}$, where $\lambda$ is a weighting parameter.

\mysection{Experiment}
Experiments were conducted to evaluate the proposed CAC transformer encoder and SDD loss function.
We constructed the AudioDiffCaps dataset consisting of pairs of similar but slightly different audio clips and a human-annotated description of their differences for the experiments.

\mysubsection{AudioDiffCaps dataset}
The constructed AudioDiffCaps dataset consists of
(i)~pairs of similar but slightly different audio clips and (ii)~human-annotated descriptions of their differences.

The pairs of audio clips were artificially synthesized by mixing foreground event sounds with background sounds taken from existing environmental sound datasets (FSD50K~\cite{fonseca2020fsd50k} and ESC-50~\cite{piczak2015dataset}) using the Scaper library for soundscape synthesis and augmentation~\cite{salamon2017scaper}. We used the same mixing procedure as our previous work~\cite{takeuchi2022introducing}.
Data labeled \myqt{rain} or \myqt{car\_passing\_by} in FSD50K was used as background, and six foreground event classes were taken from ESC-50 (i.e., data labeled \myqt{dog}, \myqt{chirping\_bird}, \myqt{thunder}, \myqt{footsteps}, \myqt{car\_horn}, and \myqt{church\_bells}).
Each created audio clip was 10 seconds long. The maximum number of events in one  audio clip was two, with 0-100\% overlap (no overlap-range control applied). 
Each foreground event class had 32 or 8 instances in the development or evaluation set, respectively. 
Similar to previous work, we focused on the three types of difference: increase/decrease of background sounds, increase/decrease of sound events, and addition/removal of sound events.
The development and evaluation sets contained 5996 and 1720 audio clip pairs, respectively. (That is, development and evaluation sets contained 11992 and 3440 audio clips.)

The human-annotated descriptions were written as instruction forms explaining "what and how" to change the first audio clip to create the second audio clip.
In the preliminary study, we found that declarative sentences, in some cases, tend to use ordinal numbers such as ``\textit{First sound} is louder than \textit{second sound}''. 
Since these cases do not express what the actual difference is, the AudioDiffCaps dataset uses instruction forms with a fixed direction of change from the first audio clip to the second one, e.g., "Make the rain louder"
\footnote{The dataset is available at \url{https://github.com/nttcslab/audio-diff-caps}.}
.
A wider variety of descriptions explaining the same concept, such as declarative sentences, should be included in future works.
The presentation order of the pair to the annotator was randomly selected.
Annotators were five na\"{i}ve workers remotely supervised by an experienced annotator.
Each pair of audio clips in the development set had between 1 and 5 descriptions (a total of 28,892) while each pair in the evaluation set had exactly five descriptions assigned to it (a total of 8600).

\begin{table*}[!t]

\caption{Results of evaluation metrics}
\label{tab:result1}
\centering
\scriptsize
\begin{tabular}{@{}C{0.07}@{}l | cc | ccccccc }
\toprule
\textbf{ID} & \textbf{System} &\textbf{Mask} &\textbf{Disent.} & \textbf{BLEU-1} & \textbf{BLEU-4} & \textbf{METEOR} & \textbf{ROUGE-L} & \textbf{CIDEr} & \textbf{SPICE} & \textbf{SPIDEr} \\	
\midrule
(a)& Baseline & N/A & N/A
& 67.1 & 31.7 & 24.3 & 56.9 & 82.7 & 19.5 & 51.1 \\
(b)& CAC transformer & Cross & N/A
& 67.0 & 33.4 & 25.2 & 59.5 & 90.2 & 19.5 & 54.9 \\

\midrule
\multicolumn{2}{l|}{\hspace{-3pt}CAC transformer}&&&&&\\
(c)& \quad w/ Early SDD  ($\lambda=0.5$) & Cross & Early
& 67.0 & 33.7 & 25.3 & 59.6 & 91.8 & 19.4 & 55.6 \\
(d)& \quad w/ Early SDD  ($\lambda=1.0$) & Cross & Early
& 66.8 & 32.2 & 25.3 & 59.3 & 91.7 & 19.5 & 55.6 \\
(e)& \quad w/ Early SDD  ($\lambda=2.0$) & Cross & Early
& 66.9 & 33.5 & 25.3 & 59.6 & 92.8 & 18.7 & 55.8 \\

&&&&&& \vspace{-3pt}\\
(f)& \quad w/ Late SDD ($\lambda=0.5$) & Cross & Late
& \textbf{70.3} & 39.2 & \textbf{26.4} & \textbf{61.6} & \textbf{97.6} & 21.3 & 59.4 \\
(g)& \quad w/ Late SDD ($\lambda=1.0$) & Cross & Late
& 69.9 & 38.3 & 26.3 & 61.5 & 96.3 & 21.2 & 58.7 \\
(h)& \quad w/ Late SDD ($\lambda=2.0$) & Cross & Late
& 69.9 & \textbf{39.5} & 26.3 & 61.3 & 97.1 & \textbf{22.6} & \textbf{59.9} \\
\bottomrule
\end{tabular}
\end{table*}

\begin{figure*}[t]
  \centering
\includegraphics[width=1.9\columnwidth]{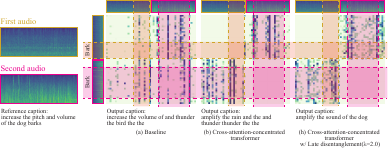} 
  \vspace{-10pt}
  \caption{
  Examples of output caption and attention weights. 
  The leftmost row was the Mel-spectrogram of two audio clips and one reference caption. 
  The three on the right were the attention weights of the transformer encoder and the output caption.
  }
  \label{fig:output_caps}

\end{figure*}

\mysubsection{Experimental conditions}
We used 10\% of the development set for validation.
The optimizer was Adam~\cite{kingma2014adam}.
The number of epochs was 100.
We used the BLEU-1, BLEU-4, METEOR, ROUGE-L, CIDEr~\cite{vedantam2015cider}, SPICE~\cite{anderson2016spice}, and SPIDEr~\cite{liu2017improved} as evaluation metrics.
They were also used for conventional audio captioning\cite{dcase2022task6}.

We used BYOL-A~\cite{niizumi2021byol-a}, a pre-trained audio embedding model, as the audio feature extractor in our ADC implementation, and we fine-tuned the BYOL-A throughout experiments.
The transformer encoder and decoder used the official implementation of PyTorch. 
The number of layers was 1. The hidden size was 768. The number of heads was 4. The activation was RELU. The dimension of the feedforward layer was 512. The dropout rate was 0.1.
For the attention mask of the transformer encoder, we compared two types; one with the proposed cross-attention mask and the other without a mask.
The text decoder used the teacher forcing algorithm during training and the beam search algorithm~\cite{koehn2009statistical, koehn2004pharaoh} during inference.
The value of $\lambda$ was empirically set to $0$, $0.5$, $1.0$, or $2.0$.

\mysubsection{Results}
The results of evaluation metrics are shown in Table~\ref{tab:result1}, where bold font indicates the highest score, ``Mask'' and ``Disent.'' indicate the attention mask utilized in the transformer encoder and input of SDD loss function, respectively.
When the CAC transformer encoder was evaluated by comparing the two lines above, the proposed method had superior or equivalent scores to the conventional method in all evaluation metrics.
There was no significant difference in the evaluation metrics related to the degree of matching with single-word references, such as BLEU-1.
One likely reason is that the scores above a certain level can be obtained by outputting words in arbitrary sentences, such as ``a'' and ``the'' in these metrics. In contrast, the scores of BLEU-4, ROUGE-L, CIDEr, and SPIDEr, affected by the accuracy of consecutive words, were improved using the proposed cross-attention mask. Therefore, the proposed cross-attention mask was thought to make the feature extraction of differences more efficient and simplify the training of the text decoder. As a result, phrase-level accuracy was improved.

The effect of SDD was verified from the results of the second to eighth lines.
The results in (a) and (b) were the conventional transformer without cross attention mask or SDD loss and the CAC transformer without SDD loss ($\lambda=0$)
Ones from (c) to (h) were the result when using early/late disentanglement.
Since the scores of BLEU-4, ROUGE-L, CIDEr, and SPIDEr improved under all conditions comparing (b) and others, the SDD loss function was effective for the audio difference captioning task.
The improvement in the case of late disentanglement (f), (g), and (h) was remarkable, and the results obtained the best scores in all evaluation metrics with late disentanglement.
In other words, it was essential to use the information to be compared to decompose the similar part and the different parts in the feature amount space. 
That corresponds to the difference determined depending on the comparison target.

Fig.~\ref{fig:output_caps} shows one of the evaluation data and estimated caption and attention weight of the transformer encoder from each system. 
The leftmost colomn is the Mel-spectrogram of the two input audio clips and one of the reference captions. 
The three on the right are the attention weight of the transformer encoder and output caption, where the attention weight shows the average of multiple heads. 
The audio clips on the left and above the weights correspond to the input and memory of the transformer, respectively. 
The area colored pink and yellow on the weights corresponds to the dog barking. 
Since there was a difference in the loudness of the dog barking between the two clips, the attention was expected to focus on areas where pink and yellow overlap to extract the difference. 

First, in (a), since the attention weight was not constrained, it was also distributed widely to areas other than the above compared with the other two. 
On the other hand, the attention weights of (b) and (h) concentrated on areas where pink and yellow overlap since the attention of the same input and memory was unavailable. 
Comparing (b) and (h), while the attention of the part containing the barking of the dog in the memory was large at any time-frame in (b), more attention was paid to the pink and yellow overlapping areas where both input and the memory contain the barking of the dog in (h). 
Since the late disentanglement required that similar and discrepant parts be retained in the output of the transformer encoder calculated using these attention weights, it was thought that the late disentanglement induced attention to be paid to the part where there was a difference when comparing the two sounds instead of paying attention to the parts that are likely to exist the difference compared with the distribution of training data, such as a dog barking.

\mysection{Conclusion}
We proposed \textit{Audio Difference Captioning} (ADC) as a new extension task of audio captioning for describing the semantic differences between similar but slightly different audio clips.
The ADC solves the problem that conventional audio captioning sometimes generates similar captions for similar but slightly different audio clips, failing to describe the difference in content.
We also propose a cross-attention-concentrated transformer encoder to extract differences by comparing a pair of audio clips and a similarity-discrepancy disentanglement to emphasize the difference feature in the latent space.
To evaluate the proposed methods, we newly built an AudioDiffCaps dataset consisting of pairs of similar but slightly different audio clips and a human-annotated description of their differences.
We experimentally showed that since the attention weights of the cross-attention-concentrated transformer encoder are restricted only to the mutual direction of the two inputs, the differences can be efficiently extracted.
Thus, the proposed method solved the ADC task effectively and improved the evaluation metric scores.

Future work includes utilizing a pre-trained generative language model such as BART~\cite{lewis-etal-2020-bart} and applying a wider variety of audio events and types of differences.

\mysection{Acknowledgments}
\vspace{-3pt}
BAOBAB Inc. supported the annotation for the dataset.

\bibliographystyle{IEEEtran}
\bibliography{refs}

\end{document}